\documentclass{ametsocV6.1}

\usepackage{amsmath,amsfonts,amssymb,bm}
\usepackage{mathptmx}
\usepackage{newtxtext}
\usepackage{newtxmath}
\usepackage{appendix}
\usepackage{enumitem}
\usepackage{graphicx}
\usepackage{subcaption}
\usepackage{lineno}
\usepackage{array}
\usepackage{longtable}
\usepackage{natbib}
\usepackage{bm}
\usepackage{tikz}
\usetikzlibrary{shapes.geometric}
\usepackage{multirow}

\newcommand{\ttau}{\mbox{ $\widetilde{\tau}$} {}}

\newcommand{\tuu}{\mbox{\boldmath $ \widetilde{u}$} {}}

\newcommand{\tp}{\mbox{$ \widetilde{p}$} {}}

\newcommand{\tu}{\mbox{$\widetilde{u}$} {}}
\newcommand{\tv}{\mbox{$\widetilde{v}$} {}}
\newcommand{\tw}{\mbox{$\widetilde{w}$} {}}
\newcommand{\thU}{\mbox{$\widehat{\widetilde{U}}$} {}}
\newcommand{\thu}{\mbox{$\widehat{\widetilde{u}}$} {}}
\newcommand{\thv}{\mbox{$\widehat{\widetilde{v}}$} {}}

\newcommand{\CDS}[1]{{{\color{black} #1}}}

\title{A sea surface-based drag model for Large Eddy Simulation of wind-wave interaction}

\authors{Aditya K. Aiyer,\aff{a}\correspondingauthor{Aditya Aiyer, aaiyer@princeton.edu} 
Luc Deike,\aff{a,b} 
Michael E. Mueller,\aff{a} 
}
\affiliation{\aff{a}{Department of Mechanical and Aerospace Engineering, Princeton University}\\
\aff{b}{High Meadows Environmental Institute, Princeton University}\\
}

\abstract{Monin-Obukhov similarity theory (MOST) is a well tested approach for specifying the fluxes when the roughness surfaces are homogeneous. For flow over waves (inhomogeneous surfaces), phase-averaged roughness length scales are often prescribed through models based on the wave characteristics and the wind speed. However, such approaches lack generalizability over different wave ages and steepnesses due to the reliance on model coefficients tuned to specific datasets. In this paper, a sea surface-based hydrodynamic drag model applicable to moving surfaces is developed to model the pressure-based surface drag felt by the wind due to the waves. The model is based on the surface gradient approach of \citet{Anderson2010} applicable to stationary obstacles and extended here to the wind-wave problem. The wave drag model proposed specifies the hydrodynamic force based on the incoming momentum flux, wave phase speed, and the surface frontal area. The drag coefficient associated with the wind-wave momentum exchange is determined based on the wave steepness. The wave drag model is used to simulate turbulent airflow above a monochromatic wave train with different wave ages and wave steepnesses. The mean velocity profiles and model form stresses are validated with available lab-scale experimental data and show good agreement across a wide range of wave steepnesses and wave ages. The drag force is correlated with the wave surface gradient and out-of-phase with the wave height distribution by a factor $\pi/2$ for the sinusoidal wave-train considered. 
These results demonstrate that the current approach is sufficiently general over a wide parameter space compared to wave phase-averaged models with a minimal increase in computational cost. 
}

\begin{document}
\nolinenumbers
\maketitle

%
%
%

%
\section*{Significance Statement}
Understanding the physics of wind-waves plays an important role in the context of numerous geophysical and engineering applications. A drag-based model is developed that characterizes the effect of the sea surface waves on the wind above. The model is validated with existing experimental datasets and is shown to be effective in predicting the average wind velocity and stress over waves with varied steepnesses and phase speeds. The ease of implementation and low computational cost of the model make it useful for studying turbulent atmospheric-scale flows over the sea surface important in offshore wind energy research as well as for modeling air-sea fluxes of momentum, heat, and mass.
\section{Introduction}
\label{sec:intro}

    The physics of wind-wave interactions involve an interplay between atmospheric turbulence and ocean waves and play an important role in the context of numerous geophysical and engineering applications.  Wind-wave interactions have been a continuous topic of research spanning  numerous experimental \citep{shemdin_hsu_1967,Grare2013,Buckley2017,Buckley2020}, theoretical \citep{Miles1993,Jannsen1991,Belcher1999,Kudryavtsev2004}, and computational \citep{Sullivan2000,Sullivan2008,Sullivan2014,Yang2010,Yang2013,Cao2020,Cao2021} studies devoted to quantifying the momentum exchange between the waves and the airflow above. Many physical processes rely on accurately quantifying the heat, mass, and momentum exchanges that take place at the air-sea interface and are important for numerous practical applications such as weather and climate modeling,  safe steering of naval vessels, and offshore wind energy harvesting.

Measurements of air pressure in the laboratory and field have been carried out to quantify the momentum exchange at the air-sea interface \citep{shemdin_hsu_1967,Snyder1981,Banner1990,Donelan2006,Peirson2008,Grare2013}. The results and findings from these experiments have been summarized by \citet{Grare2013}  where they analyzed the growth rate and associated form stress due to energy transfer between the wind and the waves for different experimental data. 
They found that the normalized wind \CDS{energy input rate} values showed good collapse as a function of wave steepness rather than the wave age that is typically used in the literature for the datasets analyzed. 
Detailed experimental measurements of the airflow structure and wind stress over the wave surface within the airflow's viscous sublayer were carried out by \citet{Buckley2017} and \citet{Buckley2020}. They found that the form stress dominated the total air-water momentum flux for high wave slopes while the viscous drag was important for low wave slopes. The high resolution and quality of the wind and wave profile measurements provide a useful database for comparison with models and simulations.

The component of the total stress $\bm{\tau}$ imparted by the wind to the wave field can be written as a sum of the pressure and tangential viscous stresses. Considering a sloped surface ${\eta}(x,t)$ varying as a function of horizontal coordinate $x$ and time $t$  representing the wave, the horizontal component of the stress can be written as\CDS{
  \begin{equation}
        \tau_{xz}   = \left\langle p_s\eta_x\right\rangle + \left\langle \frac{\tau_{visc,xz}}{\sqrt{1 + \eta_x^2}} \right\rangle,
    \end{equation}
    where $z$ is the vertical direction, $p_s$ is the surface pressure, \CDS{$\tau_{visc,xz} = \mu\partial u/\partial z$} is the local viscous surface tangential shear stress, $\eta_x = \partial {\eta}/\partial x$ is the  local interface slope and the angular brackets
denote averaging over one wavelength.}
    A more detailed discussion of the various stress components is provided in the Appendix of \citet{Grare2013}. Using a linear approximation $\eta_x <<1$, the total \CDS{energy input rate} can be related to the wave-coherent component of the horizontal stress:
\begin{equation}\label{eqn:energy_input}
    S_{in} = dE/dt = \left\langle p_s\eta_x\right\rangle c + \left\langle\tau_{visc,xz}u_s \right\rangle,
\end{equation}
where $u_s$ is the surface orbital velocity, $E = 1/2 \rho g\langle\eta^2\rangle$ is the energy per unit area of sea surface averaged over a wavelength, and $c$ is the wave phase speed. \CDS{When normalized by the wave phase speed, the first term on the RHS of Equation $\ref{eqn:energy_input}$ is defined as the form stress $\tau_{form}$ and the second term is the wave-coherent tangential stress $\tau_{visc,wc}$ \citep{Longuet-Higgins1969}.} The total energy input is then related to the total wave-coherent momentum flux as  \CDS{$S_{in} = (\tau_{form} + \tau_{visc,wc})c$.} Knowledge of the dependence of the \CDS{energy input rate} on wave characteristics is important in modeling the wind-sea momentum flux.
 Two common parameterizations for the \CDS{energy input rate} have been developed by \citet{Jeffreys1925} and \citet{miles_1957}. The former takes the form:
\begin{equation}\label{eqn:enrgy_input_jeff}
    S_{in} = \frac{1}{2} \rho_a s_{\lambda/2} (ak)^2 c^3 \left(\frac{U_{\lambda/2}}{c} - 1\right)\left|\frac{U_{\lambda/2}}{c} - 1\right|,
\end{equation}
where $\rho_a$ is the air density, $s_{\lambda/2}$ is the sheltering coefficient evaluated at a height $z = \lambda/2$ above the water surface with $\lambda$ being the wavelength, $k= 2\pi/\lambda$ is the wavenumber, $a$ is the amplitude of the wave, and $U_{\lambda/2}$ is the mean streamwise velocity at $z = \lambda/2$. The latter parameterization by \citet{miles_1957} is given by
\begin{equation}\label{eqn:enrgy_input_miles}
    S_{in} = \beta \frac{\rho_a}{\rho_w} \left(\frac{u_{\ast}}{c}\right)^2 \omega E,
\end{equation}
where $E$ is the wave energy density, $\omega$ is the angular frequency, $\rho_w$ is the water density, $u_{\ast}$ is the surface friction velocity, and $\beta$ is the normalized wave growth coefficient. The angular frequency and the wavenumber are related by the dispersion relation $\omega = \sqrt{gk}$ where $g$ is the gravitational acceleration.

Computational simulations provide an alternative approach to these theoretical parameterizations.
Direct Numerical Simulation (DNS) provides the highest fidelity description of the problem. However, these simulations have been restricted to low Reynolds number for air-flow over a wavy surface due to high computational costs associated with resolving the viscous layer for high Reynolds number \citep{Sullivan2000,lin_2008,Yang2010,Druzhinin2012}. 
Large Eddy Simulation (LES) has been shown to yield high-fidelity results for flow over complex terrain resolving the large- and intermediate-scale turbulent motions and only requiring modeling of the
unresolved turbulence effects. 
The effect of waves in LES of the marine atmospheric boundary layer (MABL)  has been incorporated either using  wave phase-resolved simulations \citep{Sullivan2008, Sullivan2014,Yang2014,Sullivan2018a,HAO2018162} or wave phase-averaged models \citep{Charnock1955,Donelan1990,Toba1990,Drennan2005}.  The former approach relies on constructing a terrain-following grid and having high enough resolution to resolve the near wave dynamics. The approach has been used successfully to simulate boundary layer flow with the bottom boundary specified using either a sinusoidal wave train \citep{Sullivan2008} or the full 3D wave spectrum \citep{Sullivan2014,Yang2014}. \citet{Yang2014} developed a dynamic coupling procedure to include the effect of air flow pressure field on the wave motion as a kinematic boundary condition. This approach is quite computationally expensive for high Reynolds number atmospheric flows due to the large separation of scales and the  high grid resolution required near the surface to  resolve the wave motion in the vertical direction.

In high Reynolds number flow simulations, a wall model can instead be used to specify the boundary fluxes \citep{Piomelli2002}. The phase-averaged approach  models the total surface stress $\tau$ by specifying a surface roughness for the waves and calculating the stress based on MOST \citep{Moeng1984}: 
\begin{equation}
    \tau = \rho_a \left[\frac{\kappa}{\log{\left(z_r/z_0\right)} - \psi_m(z/L)}\right]^2u_r^2,
\end{equation}
where $\kappa=0.4$ is the Von-Karman constant, $z_0$ is the surface roughness, $u_r$ is the reference velocity at height $z_r$, and $\psi_m$ \CDS{is an empirical function of the stability parameter,
$\zeta = -z/L$, where $L$ is the Obukhov length,} which  accounts for stratification effects. In the Monin-Obukhov similarity framework, {the unresolved stresses, including the viscous stresses, are absorbed into a single roughness length $z_0$.} The sea surface roughness elements are all in motion with waves of different wavelengths propagating with different speeds according to the wave dispersion relation. To account for this, numerous models have been proposed to parameterize an equivalent effective roughness scale $z_0$.
A widely used model is one proposed by \citet{Charnock1955} where the surface roughness  $z_0$ is calculated as
\begin{equation}\label{eqn:charnock}
    z_0 = \frac{\alpha_{ch} u_{\ast}^2}{g},
\end{equation}
 where $\alpha_{ch} = 0.01-0.03$ is an empirical constant \citep{Garatt1977}.  The wide range of values for 
 $\alpha_{ch}$ have been attributed to factors such as the wind speed, wave 
 age, wave steepness, and water depth 
 \citep{Donelan1990,Smith1992,Taylor2001,Fairall2003,Edson2013,Jimenez2018}. Alternative formulations have been proposed based on the peak wave age $c_p/u_{\ast}$ \citep{Donelan1990,Toba1990,Smith1992}:
 \begin{equation}\label{eqn:Donelan}
     z_0 = A (u_{\ast}/c_p)^{B},
 \end{equation}
 where $A$ and $B$ are empirical parameters that vary across datasets \citep{Deskos2021ReviewLayer}. 
 \citet{Taylor2001} proposed a parameterization based on the characteristic wave steepness $H_s/\lambda_p$:
 \begin{equation}\label{eqn:Taylor}
     z_0 = 1200 H_s(H_s/\lambda_p)^{3.4},
 \end{equation}
 where $H_s$ is the characteristic wave height, $\lambda_p$ is the peak wavelength, and the factors $1200$ and exponent $3.4$ are determined using data consisting of waves with larger fetch (``old waves").
 Due to the modeling uncertainties for the effective roughness in phase-averaged formulations, these formulations may not be sufficiently general. There is a need to develop a model that would be applicable to a range of wave steepnesses and wave ages yet still retaining the computational efficiency of these wave phase-averaged parameterizations. It is important to note that such a model would not resolve the detailed near-surface flow around the waves but rather reproduce the forces imposed by and flow above the waves in the atmospheric boundary layer.
 
 In the atmospheric boundary layer (ABL), the effect of forest canopies and vegetation has been represented using drag-based models termed  canopy stress models \citep{Shaw1992,Su1998,Arthur2019}. The hydrodynamic drag force due to obstacles in the flow is modeled as a function of the obstacle frontal area and the incoming momentum flux. \citet{Brown2001} applied the canopy stress models to flow over stationary sinusoidal ridges.  \citet{Anderson2010} developed a surface-gradient drag (SGD) model where the obstacle roughness was vertically unresolved but resolved horizontally. \citet{Arthur2019} implemented the canopy stress model framework in the Weather Research and Forecasting (WRF) model and applied it to both resolved and unresolved roughness.  The canopy stress approach provides a simple framework in which effects of obstacles and topographies could be included without additional computational complexity and without requiring empirical correlations for some effective roughness.

In the current work, a wave drag formulation to represent the wave stress is developed for the marine atmospheric boundary layer applicable to the wind-wave problem. The ease of implementation of this model along with the lower associated computational cost makes it an attractive alternative to more expensive wave phase-resolved terrain following approaches.
The rest of the paper is organized as follows. The governing equations used for the LES are described in  \S \ref{sec:methods}. The wave drag model formulation is described in  \S \ref{sec:wave_model}. The model validation and results are presented in \S \ref{sec:Results}. Conclusions are drawn in \S \ref{sec:Conclusions}.

\section{Computational Framework }
\label{sec:methods}
 LES calculations were performed using NGA, which is a structured, finite difference, low Mach number flow solver \citep{desjardins2008,macart2016}. The wind velocity field is described using the filtered Navier-Stokes equations in the incompressible limit:
\begin{equation}\label{eqn:div}
\nabla \cdot \tuu =0,
\end{equation}
\begin{align}\label{eqn:Navier_stokes}
\frac{\partial \tuu}{\partial t} + \tuu \cdot\nabla\tuu =& -\frac{1}{\rho}\nabla\tp- \nabla\cdot {\bm{\widetilde{\tau}}} + {\bm{f}}_d.
\end{align}
The equations are discretized on a Cartesian grid $(x,y,z)$, where $x$ and $y$ are the streamwise and spanwise coordinates, respectively, and $z$ is the vertical coordinate.
In Equations (\ref{eqn:div}) and (\ref{eqn:Navier_stokes}),  $\tuu = (\tu,\tv,\tw)$ is the velocity vector with the tilde denoting variables filtered on the LES grid; \CDS{${\bm{f}}_d$ is the force per unit mass} applied in the streamwise and spanwise directions, representing the effects of the waves; and \CDS{$\widetilde{\tau}_{ij} = -2\nu \widetilde{S}_{ij} + \ttau^d_{ij}$} is the total deviatoric stress (divided by density), where $\nu$ is the molecular viscosity, $\widetilde{S}_{ij}$ is the resolved strainrate tensor, and $\ttau^d_{ij}$ is the subfilter stress (SFS) tensor. The SFS tensor is modeled using a Lilly-Smagorinsky type subfilter viscosity model $\ttau_{ij}^d = -2\nu_T\widetilde{S}_{ij}$, where  the subfilter viscosity is computed using the Anisotropic Minimum Dissipation (AMD) model \citep{Rozema2015,Akbar2017}. The subfilter viscosity is given by
\begin{equation}
    \nu_T = \frac{-(\hat{\partial}_k \widetilde{u}_i)(\hat{\partial}_k \widetilde{u}_j)\widetilde{S}_{ij}}{(\partial_l \widetilde{u}_m)(\partial_l \widetilde{u}_m)},
\end{equation}
where $\hat{\partial}_{\gamma} = \sqrt{C}\Delta_{\gamma} \partial_{\gamma}$ (no summation implied, $\gamma = 1, 2, 3$) is the scaled gradient operator \CDS{(so dimensionless)}, {$\Delta_{\gamma}$ is the LES resolution in the respective coordinate direction,  and $C = 1/3$ is the modified Poincare constant \citep{Akbar2017}.} {The AMD subfilter model has been tested for large-scale simulations of turbulent boundary layers and shown to be accurate and have a lower computational cost compared to Lagrangian models \citep{Gadde2021}.}

In LES of high Reynolds number flows, the viscous boundary layer at the bottom boundary at the waves is unresolved. A wall-layer model is used to impose the correct surface stress for the bottom boundary. 
This surface stress could be formulated in terms of the total roughness scale $z_0$, which is generally written as a sum of the smooth surface roughness $z_{0,s}$ and a wave-based roughness scale $z_{0,w}$, giving a  total roughness $z_0 = z_{0,s} + z_{0,w}$ \citep{Yang2013}. The wave-based roughness scale is generally parameterized using phase-averaged roughness models, as discussed in the Introduction. However, in this work, the wave drag model will replace the wave-based roughness scale $z_{0,w}$ to provide better generalizability and accuracy and is discussed in the next section. For low wind speeds over smooth {ocean} surfaces, the smooth roughness scale Reynolds number, defined based on the friction velocity $u_{\ast}$ and the roughness scale $z_{0,s}$,  approaches a constant value $Re_{z_{0,s}} = 0.11$ \citep{Fairall1996}. This microscale smooth surface roughness length scale $z_{0,s}$ can then be calculated as
\begin{equation}
    z_{0,s} = 0.11\frac{\nu_a}{u_{\ast}}.
\end{equation}
To apply the resulting wall stress locally, the stress is enforced using an enhanced subfilter viscosity at the wall along with a Dirichlet boundary condition for the  velocity components.  The subfilter viscosity is set to enforce the correct wall stress given by  \citep{Moeng1984}
    \begin{equation}\label{eqn:wall_stress_x}
  \ttau_{wall} =   \left[\frac{\kappa \thU_{avg}}{\log\left(\frac{\Delta_z/2 - \widetilde{\eta}}{z_{0,s}}\right)}\right]^2,
\end{equation}
where $\thU_{avg} = \sqrt{\thu_r^2(x,y,t) + \thv_r^2(x,y,t)}$ is the magnitude of the tangential wind velocity relative to the wave surface, $ \Delta_z/2$ is the first off-wall grid point, and $\widetilde{\eta}(x,y,t)$ is the sea surface elevation filtered at the LES resolution $\Delta$. The relative surface velocities $\thu_r,\thv_r$ are calculated using the velocity at the first grid point and the wave surface orbital velocity  \citep{Yang2013,Sullivan2014,Yang2014}:
\begin{equation}
    \thu_{r} = \thu(x,y,z_1,t) - \thu_{s}(x,y,t),\qquad \thv_{r} = \thv(x,y,z_1,t) - \thv_{s}(x,y,t),
\end{equation}
where $\thu_s,\thv_s$ are the wave orbital velocities. {In the wall stress calculation, the velocities are test-filtered using a box-filter with a filter-width equal to twice the grid scale (LES resolution $\Delta$)} and are denoted by $(\widehat{\widetilde{...}})$. This additional filtering reduces velocity fluctuations significantly and improves the local applicability of the logarithmic law \citep{Bou-Zeid2005}.
The subfilter viscosity is then set such that
\begin{equation}\label{eqn:eddy_visc_wall}
    \nu_T = {\ttau_{wall}} \frac{\Delta_z/2}{\thU_{avg}} -\nu.
\end{equation}
 {A similar enhanced subfilter viscosity approach was shown to  improve predictions of statistics in the near-wall region of turbulent channel flows \citep{bae2021}}.





\section{Wave drag model}\label{sec:wave_model}
The proposed drag model for the waves is based on a canopy stress-type approach used for simulating flows over vegetation or forest canopies. The model is based on applying a hydrodynamic drag force  proportional to the incoming momentum flux onto a frontal surface area \citep{Brown2001}. This offers the advantage of incorporating the wave characteristics (surface topology and wave speed) into the formulation without relying on empirical roughness parameterizations.
Typically, for LES where the wave-air interface is unresolved, the effect of the waves is solely modeled by the subfilter stress  {via a roughness parameter}. In this section, a model is developed to represent the pressure drag due to the waves and applied locally as a force in the LES domain.

\begin{figure}
    \centering
    \includegraphics[width = \textwidth]{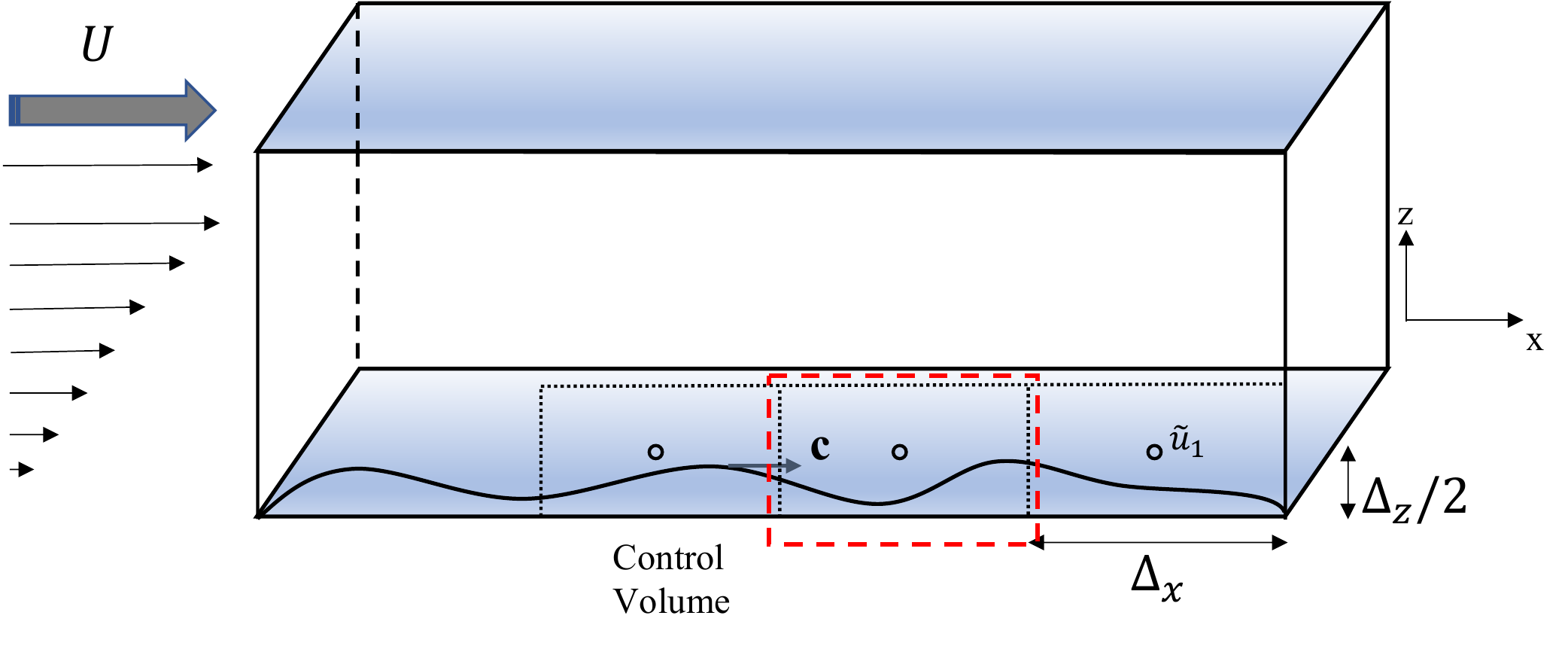}
    \caption{ Sketch of the simulation setup and the wave drag model. The waves are vertically unresolved and lie beneath the first grid point. }
    \label{fig:sketch}
\end{figure}

The surface gradient drag model of \citet{Anderson2010} is extended to account for  moving boundaries in order to model flow over waves. Figure \ref{fig:sketch} shows a sketch of the model setup. The waves are vertically unresolved and lie below the first LES grid cell similar to \citet{Anderson2010,Anderson2011}.  Consider the control volume encompassing the wave surface and the first grid point,  moving with a characteristic phase velocity $\bm{c}$. The incoming  momentum flux associated with the flow over a surface with an area $A$ is given by $\int \rho \tuu (\tuu_{r,c} \cdot \CDS{\mathrm{d}\bm{A}} )$, where $\tuu_{r,c} = \tuu - \CDS{\bm{e_x}c}$ is the 
relative velocity between the incoming flow and the wave phase velocity, where \CDS{$\bm{e_x}$ is the unit vector in the $x$-direction} and $\bm{\widetilde{u}}$ are the velocities evaluated at $z = \Delta_z /2$.
It is important to note that the phase velocity has been used in the current formulation to preserve Galilean invariance.   {The model can be generalized to waves represented by a wave spectrum. Each wave component imposes a drag proportional to the relative velocity  $\widetilde{\bm{u}} - \bm{e_x}{c}$  with the total drag calculated as a summation of individual effects.} This formulation ensures that the drag for waves with different wavelengths propagating with different wave speeds is incorporated in the formulation, with the force contribution from each wavelength calculated separately.

The flow frontal area normal to the incoming flow can be calculated based on the surface gradient as $\bm{\hat{n}_u}\cdot \nabla \eta \Delta_x \Delta_y$, where 
$\bm{\hat{n}_u} = \tuu_{r,c}/U^{\Delta}$ and 
$U^{\Delta} = \sqrt{(\widetilde{u} - c_x)^2 + (\widetilde{v}-c_y)^2}$ is the tangential velocity relative to the wave propagation speed.  The momentum flux can then be written as $M_{wave} = \rho \tuu U^{\Delta} \bm{\widehat{n}_u}\cdot \nabla \widetilde{\eta} \Delta_x \Delta_y$. In the wind-wave problem, only a fraction of the incoming momentum flux is imparted as drag and quantified in the model through a drag coefficient $C_D$. The wave drag forcing term is included  in the momentum equation by dividing the momentum flux by the local grid volume. Additionally, the forcing is only applied if $\bm{\widehat{n}\cdot \nabla \widetilde{\eta}} >0$, that is, only on the "front" of the wave with respect to the flow.
The wave drag model is then expressed as
\begin{equation}\label{eqn:drag_eqn}
    F_{d,i} = - \CDS{C_D}\frac{M_{wave}}{\Delta_x\Delta_y\Delta_z}= -C_D\frac{\rho}{\Delta_z}\widetilde{u}_i U^{\Delta} \left(\widehat{n}_{u,k}\cdot \frac{\partial \widetilde{\eta}}{\partial x_k}\right) {H\left\{\widehat{n}_{u,k}\cdot \frac{\partial \widetilde{\eta}}{\partial x_k}\right\}}\qquad i =x,y,
\end{equation}
 where $H[x]$ is the Heaviside function that ensures that the drag force is only applied when the flow is incident on the wave frontal area. 
 The wave surface height is not resolved on the LES grid, and its effects are completely represented by Equation (\ref{eqn:drag_eqn}). However, the waves must be well resolved in the horizontal direction to ensure adequate representation of the wave topology, that is, surface gradients of $\widetilde{\eta}$.

The drag coefficient can be determined by considering the \CDS{energy input rate} from the the wind to the waves as described in Equations (\ref{eqn:enrgy_input_jeff}) and (\ref{eqn:enrgy_input_miles}). The form derived by \citet{Jeffreys1925} gives a quadratic dependence of the \CDS{energy input rate} for both the wave steepness $ak$ and the inverse wave age $c/U_{\lambda/2}$ (or equivalently $c/u_{\ast}$).
The \CDS{energy input rate} from the model for a sinusoidal wave train can be calculated by substituting $\widetilde{\eta} = a\cos{(k(x-ct))}$ in Equation (\ref{eqn:drag_eqn}):
\begin{equation}\label{eqn:energy_input_les}
     S_{in,LES} = F_d\Delta_z c  =  C_D \rho (ak) c^3 \frac{\tu_{\Delta_z/2}}{c} \left(\frac{\tu_{\Delta_z/2}}{c}-1\right).
\end{equation}
Note that the \CDS{energy input rate} is evaluated by integrating the Heaviside function over one wavelength and assuming that the velocity at the first grid point is approximately constant and any resulting prefactors are absorbed into $C_D$. The model \CDS{energy input rate} varies linearly with the wave steepness and  quadratically with wave age. Comparing Equations (\ref{eqn:enrgy_input_jeff}) and (\ref{eqn:energy_input_les}), the drag coefficient can be set proportional to the wave steepness $C_D = C_{\rm LES}\cdot ak$ to recover a scaling similar to the
\citet{Jeffreys1925} parameterization for the wave \CDS{energy input rate}. $C_{\rm LES}$ should be an order unity coefficient assuming the sheltering coefficient to be a constant.
 \citet{Peirson2008} analyzed the dependence of the sheltering coefficient on the wave steepness and found that the values were constant for $ak > 0.1$. For lower wave steepness $ak < 0.1$, the sheltering coefficients were higher. In order to mimic the higher sheltering coefficients for lower wave steepness, the constant $C_{\rm LES}$ can be adjusted to weakly depend on the wave steepness. This allows for modulation of the drag coefficient for the low and high $ak$ limits. For instance, \citet{Belcher1999} developed a theoretical framework that incorporated a wave coherent tangential stress contribution as well as a modification to the turbulent stress adjacent to the interface. This yielded an expression for the form stress that was proportional to $(ak)^2$ but modulated by a prefactor $C_{Belcher} = (\beta_f + \beta_t)/(2 + \beta_f(ak)^2)$, where $\beta_f$ and $\beta_t$ are contributions due to the form and tangential stress respectively. For values of $\beta_f = 12$ and $\beta_t = 20$ \citep{Peirson2008}, the prefactor varies from $C_{Belcher} = 16$ (equal to that derived by \citet{Plant1982}) at $ak = 0$ to $C_{Belcher} = 11$ at $ak = 0.27$.  Incorporating a similar behavior into $C_{\rm LES}$, the drag coefficient $C_D$ can be written as
\begin{equation}
    C_D = \frac{P}{1 + Q(ak)^2} ak,
\end{equation}
where  $P = 1.2$ and $Q = 6$ to be consistent with \citet{Belcher1999}.
The drag force per volume applied in LES at the first grid point can now be written as
\begin{equation}\label{eqn:drag_eqn_final}
    {F}_{d,i} = -\frac{P}{1 + Q(ak)^2}(ak)\frac{\rho}{\Delta_z}\widetilde{u}_i U^{\Delta} \left(\widehat{n}_{u,k}\cdot \frac{\partial \widetilde{\eta}}{\partial x_k}\right) {H\left\{\widehat{n}_{u,k}\cdot \frac{\partial \widetilde{\eta}}{\partial x_k}\right\}} \qquad i = x,y.
\end{equation}
 The force acts only in the horizontal directions, to remain consistent with the MOST implementation (Equation (\ref{eqn:wall_stress_x})). This assumption neglects the effects of a surface-normal velocity  that could become important for swell conditions.
 It is important to note that Equation (\ref{eqn:drag_eqn_final}) is written for moderate wave phase speeds or ``young-waves”. In the case of swell, where the wave age is much higher, momentum is transported from the waves to the wind. The efficiency of momentum transfer from the waves to wind would modify the drag coefficient used in the model and is left for future studies. 

 {
The wave drag model applies a body force calculated based on the wave characteristics at the first-grid point of the simulation. For vertically unresolved surface waves, this force can be recast as a stress and applied as a boundary condition. This formulation is equivalent to the body force formulation. Casting the wave drag as a body force allows intuitive extension of the current model to other problems of interest including  wind-wave misalignment and floating wind turbines where the hydrodynamic wave force plays an important role. An equivalent surface stress and total roughness formulation are presented here for the sake of completeness.}
\begin{enumerate}[topsep = 0ex, parsep=0ex, partopsep = 0ex]
 {\item Surface stress approach
 
 The total stress at the wall can be written as a sum of the contribution from the Monin-Obukhov similarity and the wave drag model:
\begin{equation}\label{eqn:stress_part}
    \widetilde{\tau}_{iz} = \widetilde{\tau}_{iz}\vert_{wall}(x,y) + \frac{1}{\rho}{F}_{d,i}(x,y,\Delta_z/2)\Delta_z  \qquad i =x,y.
\end{equation}
The first term on the right hand side represents the portion of the stress modeled due to the smooth roughness, and the second term represents the stress due to the resolved wave field calculated using the wave drag model.
    \item  Equivalent roughness approach
    
    These boundary conditions can also be collectively applied using the standard logarithmic formulation using a total roughness including the waves, with a spatially and temporally evolving wave roughness length.
Using equations (\ref{eqn:wall_stress_x}) and (\ref{eqn:drag_eqn_final}) in (\ref{eqn:stress_part}) and rewriting for the total stress, the total stress is given by

\begin{align}
    \tau_{iz}^T  &= \left(\left[\frac{\kappa}{\log\left(\frac{\Delta_z/2 - \eta}{z_{0,s}}\right)}\right]^2  \thu_{r,i}\thU_{avg}  + C_D  \left(\widehat{n}_{u,k}\cdot \frac{\partial \widetilde{\eta}}{\partial x_k}\right) {H\left\{\widehat{n}_{u,k}\cdot \frac{\partial \widetilde{\eta}}{\partial x_k}\right\}}{\widetilde{u}_i U^{\Delta}}\right).
\end{align}

The total roughness then can be calculated as:

\begin{equation}\label{eqn:z0_WDM}
    z_{0}^{WDM} = \left(\frac{\Delta_z}{2} -\eta\right)\exp{\left[-\kappa\left(\left(\frac{\kappa }{\log\left(\frac{\Delta_z/2 - \eta}{z_{0,s}}\right)}\right)^2  + C_D \left(\widehat{n}_{u,k}\cdot \frac{\partial \widetilde{\eta}}{\partial x_k}\right) {H\left\{\widehat{n}_{u,k}\cdot \frac{\partial \widetilde{\eta}}{\partial x_k}\right\}}\right)^{-\frac{1}{2}}\frac{\widetilde{u}_i U^{\Delta}}{\thu_{r,i}\thU_{avg}}\right]}.
\end{equation}

This roughness can then be applied in the simulation to parameterize the effects of the waves. The wave component of the total roughness depends on the relative velocity of the airflow to the wave surface and implicitly depends on the velocity at the first grid point in the LES domain. In the limit of very slow waves $\left({\widetilde{u}_i U^{\Delta}}\right)/\left(\thu_{r,i}\thU_{avg}\right)\approx 1$, a phase-averaged roughness length can be calculated using Equation \ref{eqn:z0_WDM} assuming the direction of wind propagation is known.}
\end{enumerate}
\section{LES of wind over a sinusoidal wave-train}\label{sec:Results}
\subsection{Simulation setup}
 The wave surface is specified by $\widetilde{\eta}(x,t) = a \cos\left({k\left(x - ct\right)}\right)$, and
the orbital velocities are defined using the deep water wave solutions:
\begin{equation}
    u_s = a\omega\cos(k(x-ct)) ;\qquad v_s = 0 ; \qquad w_s  = a\omega\sin{(k(x-ct))}.
\end{equation}
The computational domain size in the horizontal directions is $L_x=L_y = 5\lambda$. The horizontal grid is discretized using sufficient points in the streamwise and spanwise directions, ensuring that the wave is horizontally well-resolved on the LES grid with \CDS{the grid resolution for each case provided in Table 1.}  {The vertical extent is of the order of a wavelength in order to facilitate comparison with the experiments.} The discretization $N_z$ is chosen to ensure that the wave lies below the cell center of the first grid point, $a  = 0.95\Delta_z/2$, and  \CDS{that the grid aspect ratio is ideal for wall-modeled LES, i.e $\Delta_x/\Delta_z \geq 1$ \citep{Piomelli2002,kawai2012}. The vertical resolution $\Delta_z = 0.02-0.06 H$, where $H$ is the height of the boundary layer, is a typical resolution for wall-modeled LES \citep{Nicoud2001,Piomelli2002,kawai2012,Lee2013,bae2021}.} Therefore, the wave is unresolved in the vertical direction, and its effects are solely specified by the wave drag model described in the previous section. Details of the grid and wave properties for the different configurations are provided in Table \ref{tab:Sim_param}.  Periodic boundary conditions are applied in the $x$ and $y$ directions, and the flow is driven with a constant external pressure gradient. The pressure gradient results in a friction velocity  (or surface stress) $ u_{\ast}^2 =\rho^{-1}(\partial P/\partial x) H$. A free-slip boundary condition is used for the top of the domain. The simulations are run for a total 200  {eddy turnover times} $T_{eddy} = H/u_{\ast}$ with averaging carried out over the last 50  {turnover} times. The simulation setup is depicted in Figure 1.   {The height $H$ can be calculated from the values provided in Table 1 as the wavelength $\lambda$ can be determined using $\lambda = 2\pi/k$ where $k$ is the wavenumber, calculated using the relation $ k = g/c^2$.}
 \CDS{
 \begin{table}
\centering
\begin{tabular}{c c c c c c c} 
 Case & 
 \centering $ak$
 & $c/u_{\ast}$ & $u_{\ast}$ & $N_x$& $N_z$ & $H/\lambda$ \\ [5pt] \hline
 \multirow{3}{*}{\citet{Buckley2020}}  & $0.06$& $6.57$ &$0.073$ &$128$ & 48 & 1 \\ 

 &$0.12$  &  $3.91$ & $0.167$ & $64$ & $28$& 1 \\ 
 &$0.2$  & $1.8$ & $0.538$ &$64$ & $32$ & 2 \\
{\citet{Buckley2016,Buckley2017}} &$0.27$ & $1.4$ & $0.672$ & $48$ &$24$ & $2$ \\
 \hline
 \multirow{3}{*}{Different Wave Steepnesses}& 0.1& $4.2$ & $0.2$ &$64$ & $16$ & 1  \\ 
 &$0.15$  &  $4.2$ & $0.2$ & $64$ & $16$ & 1  \\ 
 &$0.2$  & $4.2$ & $0.2$ & $64$ &$16$ & 1 \\
   \hline
\multirow{2}{*}{Different Wave Ages}
 &$0.1$  &  $7.84$ & $0.111$& $64$ & $16$ & 1  \\ 
 &$0.1$  & $11.6$ & $0.076$ & $64$ &$16$ & 1 \\
 
\end{tabular}
\caption{Wave train simulation parameters.}
\label{tab:Sim_param}
\end{table}
}
\subsection{Description of datasets}

The model is validated using laboratory-scale experiments from \citet{Buckley2016,Buckley2017,Buckley2019} and \citet{Buckley2020}. The validation datasets consist of a wide parameter space with wave steepnesses ranging from $ak = 0.06$ to $ak = 0.27$ and wave ages from $c/u_{\ast} =1.4$ to $c/u_{\ast} = 7$.

\citet{Buckley2020} carried out two-dimensional high resolution measurements of the airflow above wind-generated waves. The experiment consisted of a tank specifically designed for the study of air-sea interactions where wind-waves were generated by a computer-controlled recirculating wind tunnel. The apparatus and data processing techniques are described in detail in \citet{Buckley2017}. A total of four cases are presented combining data from \citet{Buckley2017} and \citet{Buckley2020}. The wave characteristics of the cases considered are provided in Table \ref{tab:Sim_param}.  {The wave characteristics are chosen based on the peak frequencies from the experiments. The experimental waves were generated using a wave paddle forced at a particular frequency resulting in a narrow-banded spectrum.}


\subsection{Results and discussions}

\subsubsection{Velocity Statistics and Shear Stress}
In this section, the emphasis is on the  {instantaneous} and mean velocity profiles obtained using the wave drag model for different wave steepnesses and wave ages.

\begin{figure}
    \centering
    \includegraphics[width=1\textwidth]{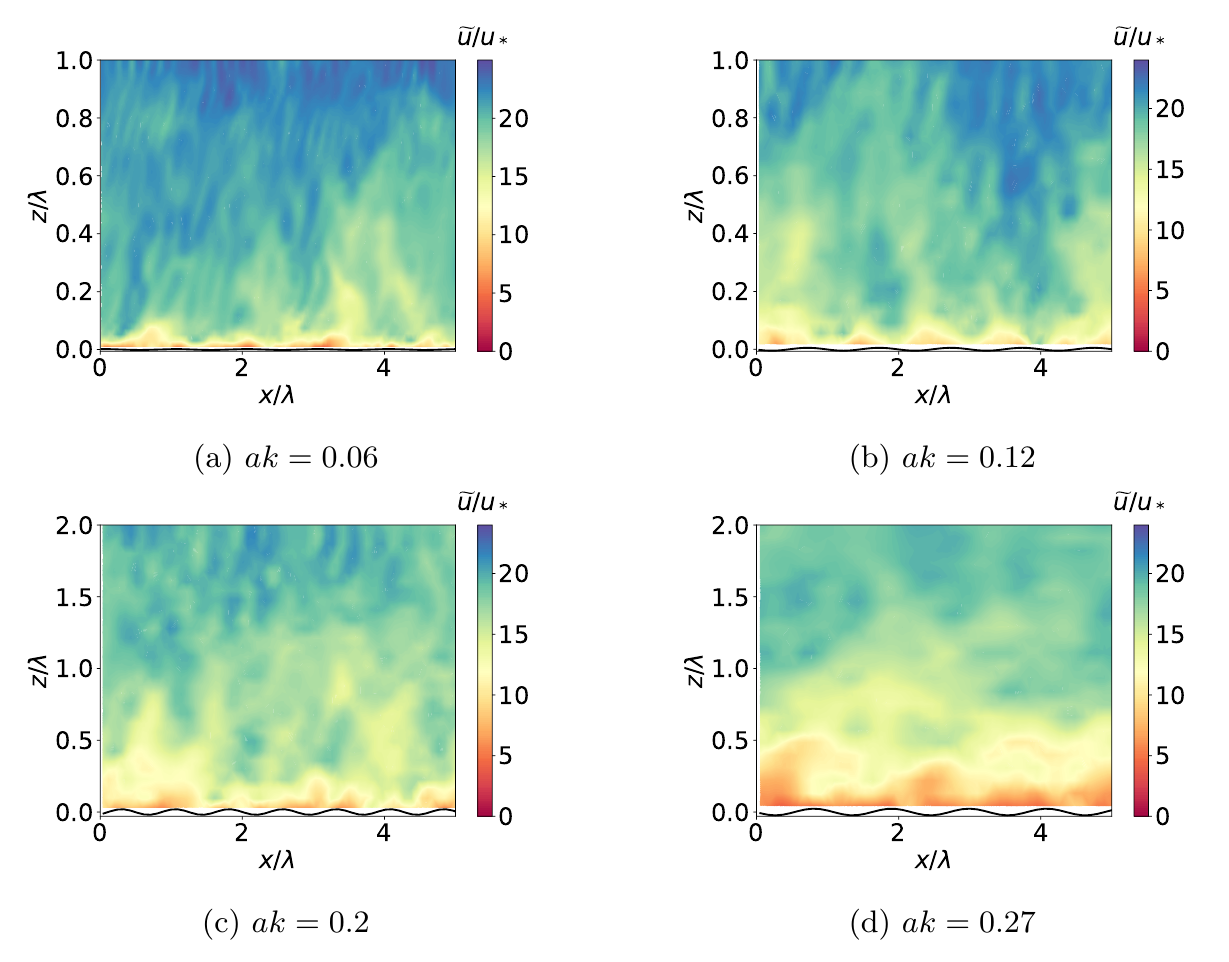}
    \caption{Instantaneous velocity profiles from LES-WDM 
     for a) $ak = 0.06$, b) $ak = 0.12$, c) $ak = 0.21$, and 
     \mbox{d) $ak = 0.27$}. Additionally shown are the wave height distribution in the black line. The effect of the wave phase is captured in the instantaneous velocity profiles.}
    \label{fig:inst_vel}
\end{figure}
 {The $x-z$ contours at the spanwise mid-plane of the instantaneous streamwise velocity are shown in Figure \ref{fig:inst_vel}. A reduced value of streamwise velocity is observed at the crests in each case. The decrease in streamwise velocity extends to a larger vertical extent for the cases with higher steepnesses. Figure \ref{fig:inst_vel}a depicts the case with the lowest steepness and is closest to a rough wall turbulent boundary layer with mild phase effects. For the steepest wave case, the phase effects extend to about 20\% of the boundary layer height. The ``phase-aware'' nature of the model is well captured in the instantaneous velocity contours.}
 
\begin{figure}
    \centering
    {\includegraphics[width=0.55\textwidth]{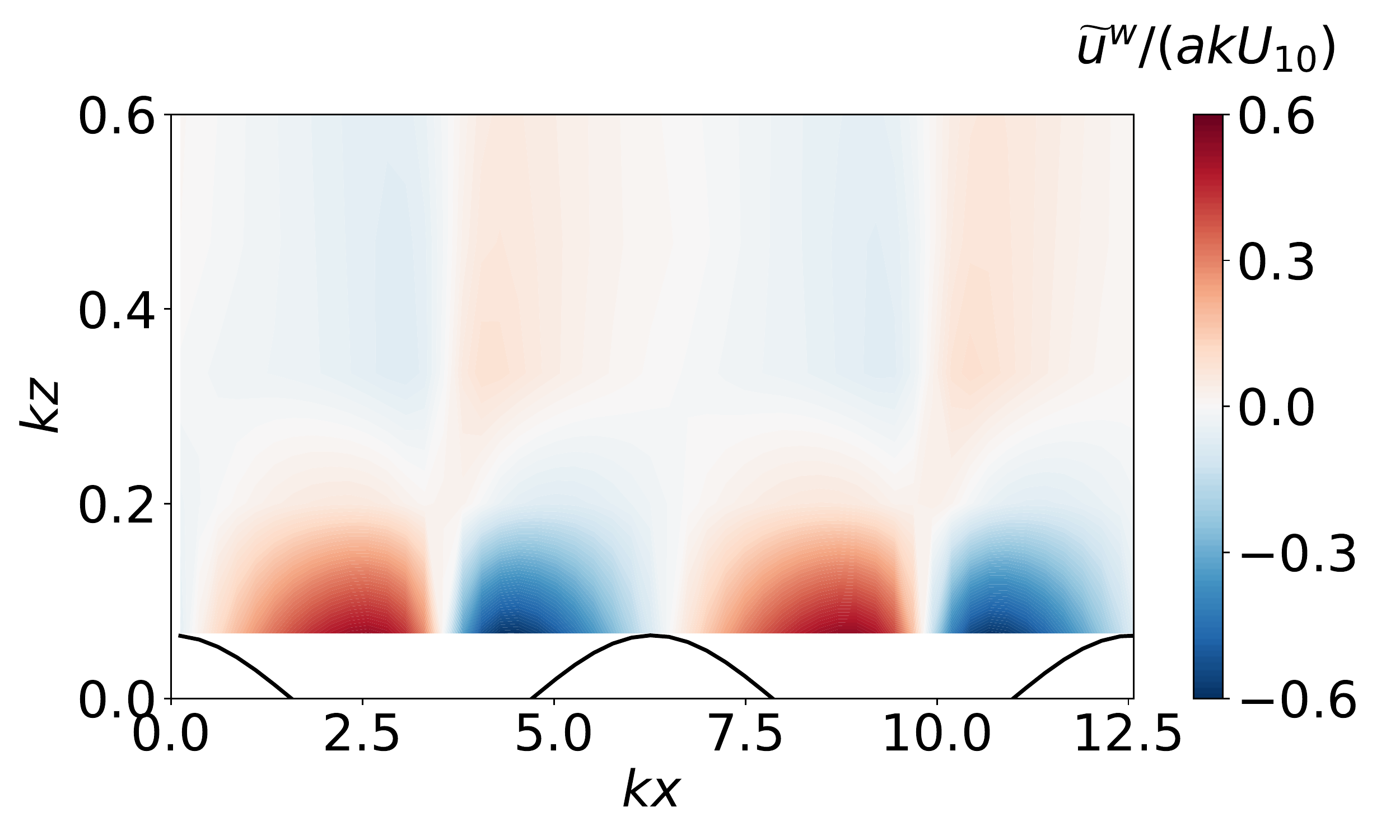}}
    \caption{\CDS{ Contours of the normalized streamwise wave fluctuation velocity $\tilde{u}^w$ near the wave surface for $ak = 0.06$ and $U_{10} = 2.19$ m/s. At the first gridpoint above the wave surface the velocity decelerates as the flow encounters the wave crest (due to the applied drag force) and accelerates downwind. The phase aware nature of the model manifests in the form of the alternate positive and negative wave fluctuation velocities.}}
    \label{fig:wave_fluc}
\end{figure}

\CDS{To highlight the wave-coherent streamwise motions, a triple decomposition of the velocity field is used (\textit{Hara and Sullivan (2015)}). The field is first split into an ensemble (Reynolds) average and a turbulent fluctuation, $\phi = \overline{\phi} + \phi^{\prime}$, where the prime denotes the turbulent fluctuation. The ensemble average $\overline{\phi}$ is further split into the horizontal mean $\langle\phi\rangle$ and a wave fluctuation ${\phi^w}$. A triple decomposition of the velocity field can then be written as 
 \begin{equation}
 \widetilde{u}  = \overline{ \tilde{u}} +  \tilde{u}^{\prime} = \langle \tilde{u}\rangle  +\tilde{u}^w + \tilde{u}^{\prime}.
 \end{equation}
Figure \ref{fig:wave_fluc} shows the normalized streamwise wave fluctuation velocity contours near the wave surface for $ak=0.06$. The streamwise velocity fluctuation from the model shows good qualitative agreement with the experiments of \citet[see Figure 4a]{Buckley2019} above $kz = 0.2$. At the wave surface there are discrepancies since the current model does not include fluctuations in the velocity due to the vertical wave motions.}

\begin{figure}
    \centering
    \includegraphics[width=1\textwidth]{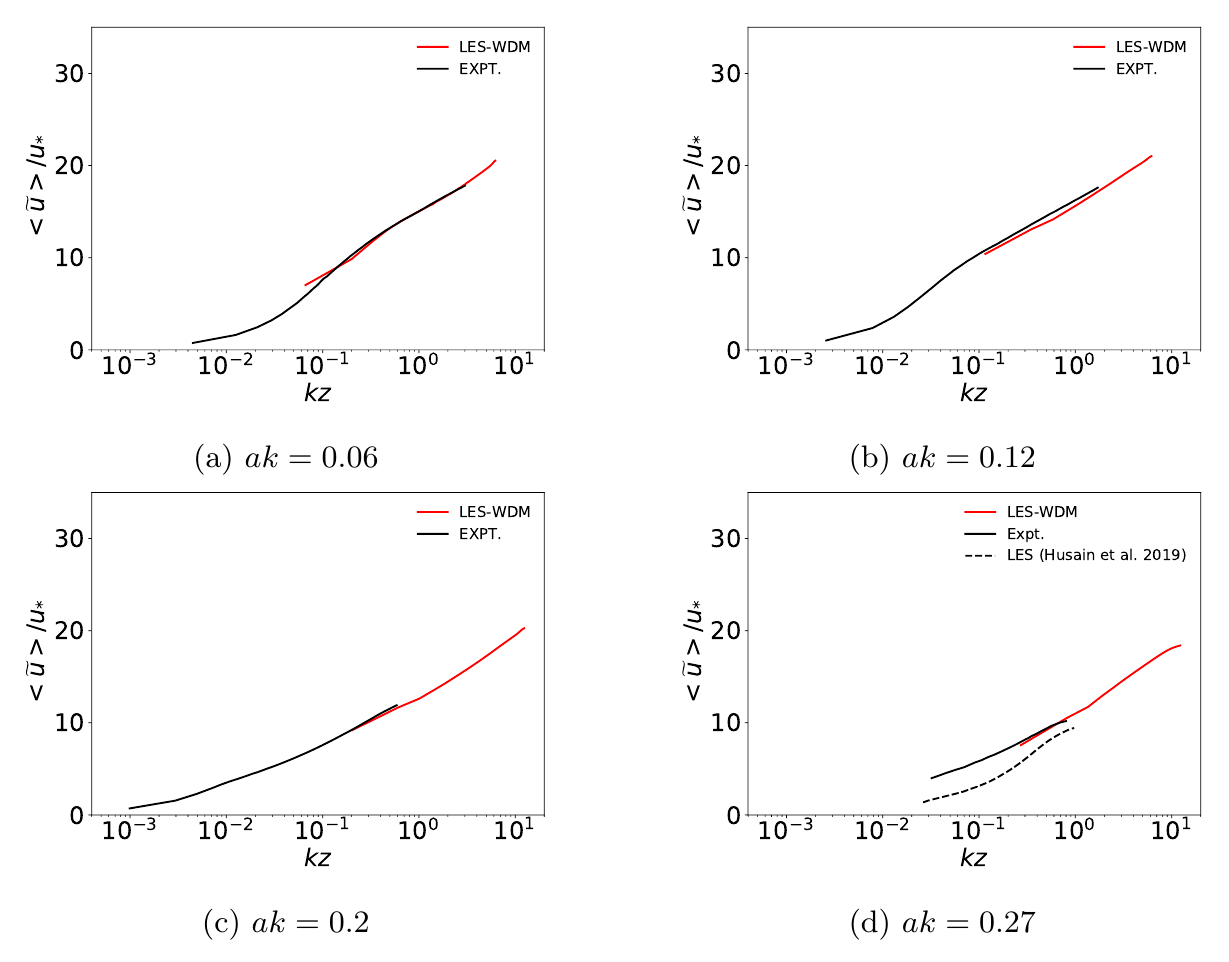}
 \caption{Mean velocity profiles from LES-WDM 
    and experiments of \citet{Buckley2020} 
     for a) $ak = 0.06$, b) $ak = 0.12$, c) $ak = 0.21$, and 
    \mbox{d) $ak = 0.27$}. 
    The wave phase-resolved LES from \citet{Husain2019} 
    is shown in d). }
    \label{fig:Buckley_2020}
\end{figure}

In Figure \ref{fig:Buckley_2020}, the mean velocity profiles are shown from the LES with the wave drag model, hereafter referred to as LES-WDM, for the different wave steepnesses and compared to the experimental data. As the wave steepness increases and the wave age decreases, the mean wind velocity measured by the experiments is reduced due to increasing resistance from the waves. The LES-WDM accurately captures this effect with slower velocities predicted for higher wave steepness and low wave age.

\begin{figure}
    \centering
    \includegraphics[width=\textwidth]{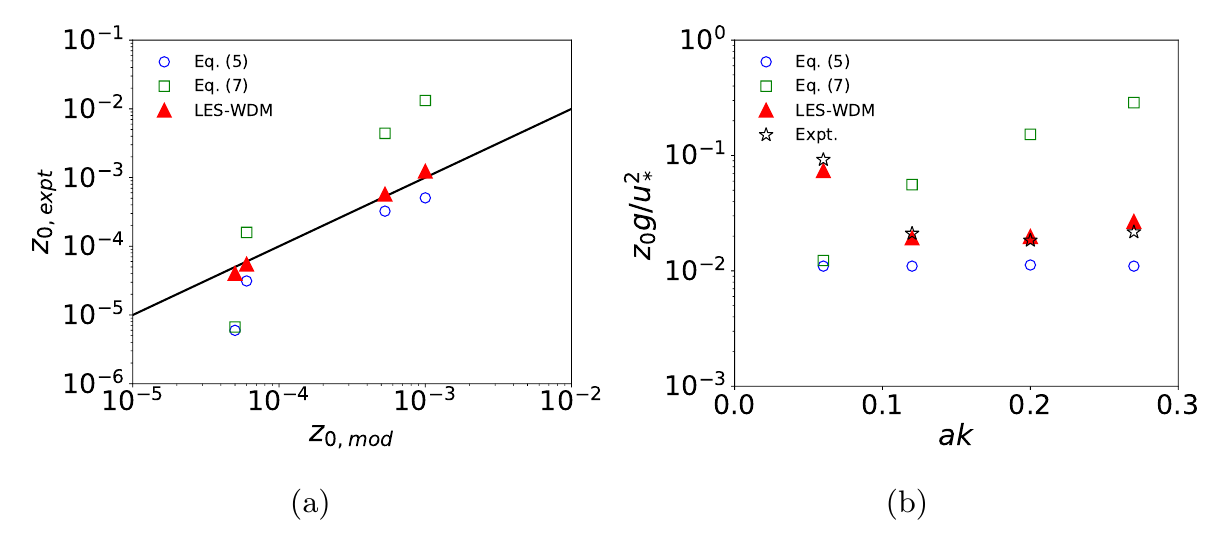}
    \caption{Total roughness length calculated from the experimental data and roughness models (Eqs. (\ref{eqn:charnock}), (\ref{eqn:Taylor}) and (\ref{eqn:z0_WDM})): a) parity plot of experimental and modeled phase-averaged roughness length and b) variation of normalized total roughness with wave steepness. The solid line in the left plane is where $z_{0,expt} = z_{0,mod}$.}
    \label{fig:z0_plots}
\end{figure}

Figure  \ref{fig:Buckley_2020}a shows the comparison for $ak=0.06$. For small wave steepnesses the deviation from the log-law is small, with $\kappa = 0.4$.  {The LES-WDM generalizes across the different experimental conditions and shows good agreement for all cases considered. The effective roughness lengths from the experiment and the models are compared in Table \ref{tab:z0s}. The roughness length for the experimental data is calculated by fitting a logarithmic profile to the data. The roughness length from the simulation can be calculated similarly or by averaging equation Equation (\ref{eqn:z0_WDM}) and using the velocity profile from the LES. Additionally, for each case, the roughness lengths obtained by using the Charnock model and the steepness-dependent roughness model, calculated using Equations (\ref{eqn:charnock}) and (\ref{eqn:Taylor}), respectively, are shown in Table \ref{tab:z0s}.  Figure \ref{fig:z0_plots}a compares the modeled experimental 
results of Table \ref{tab:z0s} in the form of a parity plot. The total roughness, in general, can depend on the wind speed (or friction velocity $u_*$), the wave age $c/u_*$, and/or the wave steepness $ak$. In the experimental dataset considered, all three parameters are
changing for the different cases. An effective Charnock parameter can be calculated by normalizing the total surface roughness by $\alpha_{ch,eff} = u_*^2/g$ and is shown as a function of wave steepness in Figure \ref{fig:z0_plots}b. The normalized roughness for the Charnock model is a constant $\alpha_{ch} = 0.11$ as the length scale predicted by the Charnock model only depends on the friction velocity and has no explicit dependence on the wave steepness or wave age. In the Taylor model, the normalized roughness increases as a function of wave steepness. It is evident from the experimental data that the effect of the waves are more complicated as the normalized roughness depends on both the wave age and the steepness. This trend is also captured in the current LES-WDM.} Modeling the form drag using the LES-WDM provides remarkable improvement over the
standard phase-averaged approaches (Eq. \ref{eqn:charnock} \& \ref{eqn:Taylor}) as it incorporates the effect of the wind speed, the wave age, and the wave steepness. The effective roughness length scale predicted by the phase-averaged model of \citet{Charnock1955} is smaller than the corresponding experimental roughness, while the roughness length calculated using the model proposed by \citet{Taylor2001} is larger.  The steepness dependent model only accounts for the wave steepness effect, neglecting the wave propagation speed. The results from the Charnock model could be improved by using a higher value of the parameter $\alpha_{ch}$; however, this constant cannot be determined \emph{a priori} and would vary with the other wave characteristics.

Note that the LES-WDM formulation is developed for coarse grids and the focus is on predicting the velocity profiles above the peak wave amplitude. Therefore, for the higher wave steepnesses only a few points lie in the range of the experimental data as the near-field of the wave is unresolved.

 \begin{table}
\centering
\begin{tabular}{c c c c c c } 
 $ak$
 & $c/u_{\ast}$ & $z_{0,Expt.}$ & $z_{0,LES}$ & $z_{0,Charnock}$ & $z_{0,Taylor}$ \\ [5pt] \hline
 
 $0.06$& $6.57$ &$5.5\times 10^{-5}$ & $6\times 10^{-5}$ & $8.15\times 10^{-6}$  & $6.69\times 10^{-6}$\\ 

 $0.12$  &  $3.91$ &$6.0\times 10^{-5}$ & $7\times 10^{-5}$ & $4.26\times 10^{-5}$  & $1.60\times 10^{-4}$\\ 
 
 $0.2$  & $1.8$ &$2.3\times 10^{-4}$ & $4\times 10^{-4}$ & $4.42\times 10^{-4}$  & $4.43\times 10^{-3}$ \\
$0.27$ & $1.4$ &$1.0\times 10^{-3}$ & $1.4\times 10^{-3}$ & $6.91\times 10^{-4}$  & $1.32\times 10^{-2}$ \\

\end{tabular}
\caption{Total roughness length calculated from the experimental data, the phase-averaged roughness models, and the LES-WDM for the different cases.}
\label{tab:z0s}
\end{table}

\begin{figure}
    \centering
    \includegraphics[width=1\textwidth]{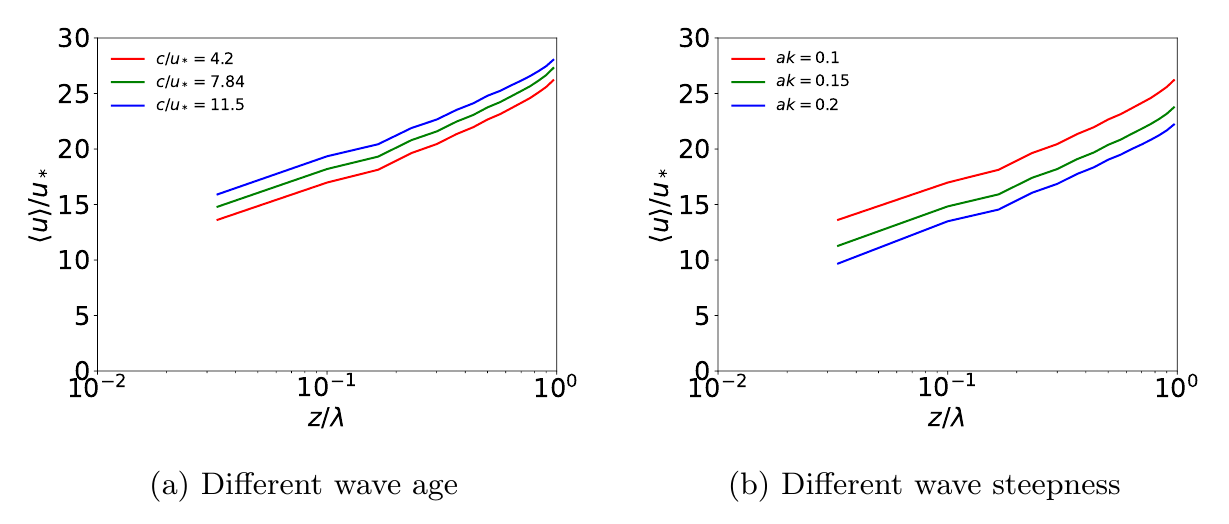}
    \caption{Normalized mean velocity profiles for  a) different wave ages $c/u_{\ast} = 4.2,\ 7.84$, and $11.5$ with wave steepness fixed at $ak = 0.1$  and  b) different wave steepnesses $ak = 0.1$, $ak = 0.15$, and $ak = 0.2$ and wave age fixed at  $c/u_* = 4.2$. 
     The mean velocity is faster for higher wave ages and lower wave steepnesses.}
    \label{fig:DNS_Sullivan_comp}
\end{figure}

To highlight and isolate the effects of wave age and steepness on the airflow, two additional sets of simulations are considered. First, the wave steepness is fixed at $ak =0.1$, and the wave age varies as $c/u_* = 4.2$, $c/u_* = 7.84$, and $c/u_* = 11.5$. The second set of simulations has fixed wave age $c/u_* = 4.2$ and wave steepnesses $ak = 0.1$, $ak = 0.15$, and $ak = 0.2$. Figure \ref{fig:DNS_Sullivan_comp}a shows that the normalized mean velocity is faster for higher wave ages. Based on Equation \ref{eqn:drag_eqn_final}, increasing the wave phase speed reduces the wave form drag applied by the model. The wave offers smaller resistance to the wind, and the blocking effect on the airflow is diminished. In Figure \ref{fig:DNS_Sullivan_comp}b the effect of only changing the wave steepness is highlighted. Larger steepness of the wave results in higher blockage thereby applying more drag on the airflow. Additionally, the reduction in the overall mean velocity profile is not uniform between $ak=0.1$, $ak = 0.15$ and $ak = 0.2$ as the effect of the steepness on the wave drag is not linear.

\begin{figure}
    \centering
    \includegraphics[width=\textwidth]{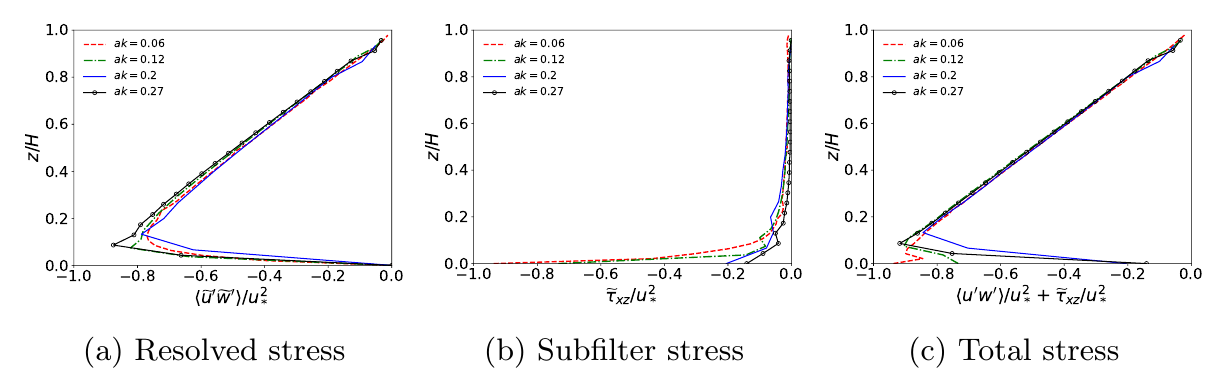}
    \caption{Mean normalized a) resolved stress, b) subfilter stress, and c) total   shear stress profiles for $ak = 0.06$, $c/u_{\ast} = 6.57$; 
    $ak = 0.12$, $c/u_{\ast} = 3.91$; 
    $ak = 0.2$, $c/u_{\ast} = 1.8$; 
    and $ak = 0.27$, $c/u_{\ast} = 1.4$. 
    The resolved stress peaks just above the surface for each case. The magnitude of the normalized subfilter surface stress is less than the imposed stress, $|\ttau_{xz}\vert_{wall}| <1$, with the remaining contribution from the wave drag model. Note $H=\lambda$ for $ak=0.06$ and $ak=0.12$, and $H=2\lambda$ for $ak=0.2$ and $ak=0.27$.}
    \label{fig:uv_LES}
\end{figure}

\begin{figure}
    \centering
  \includegraphics[width = 0.6\textwidth]{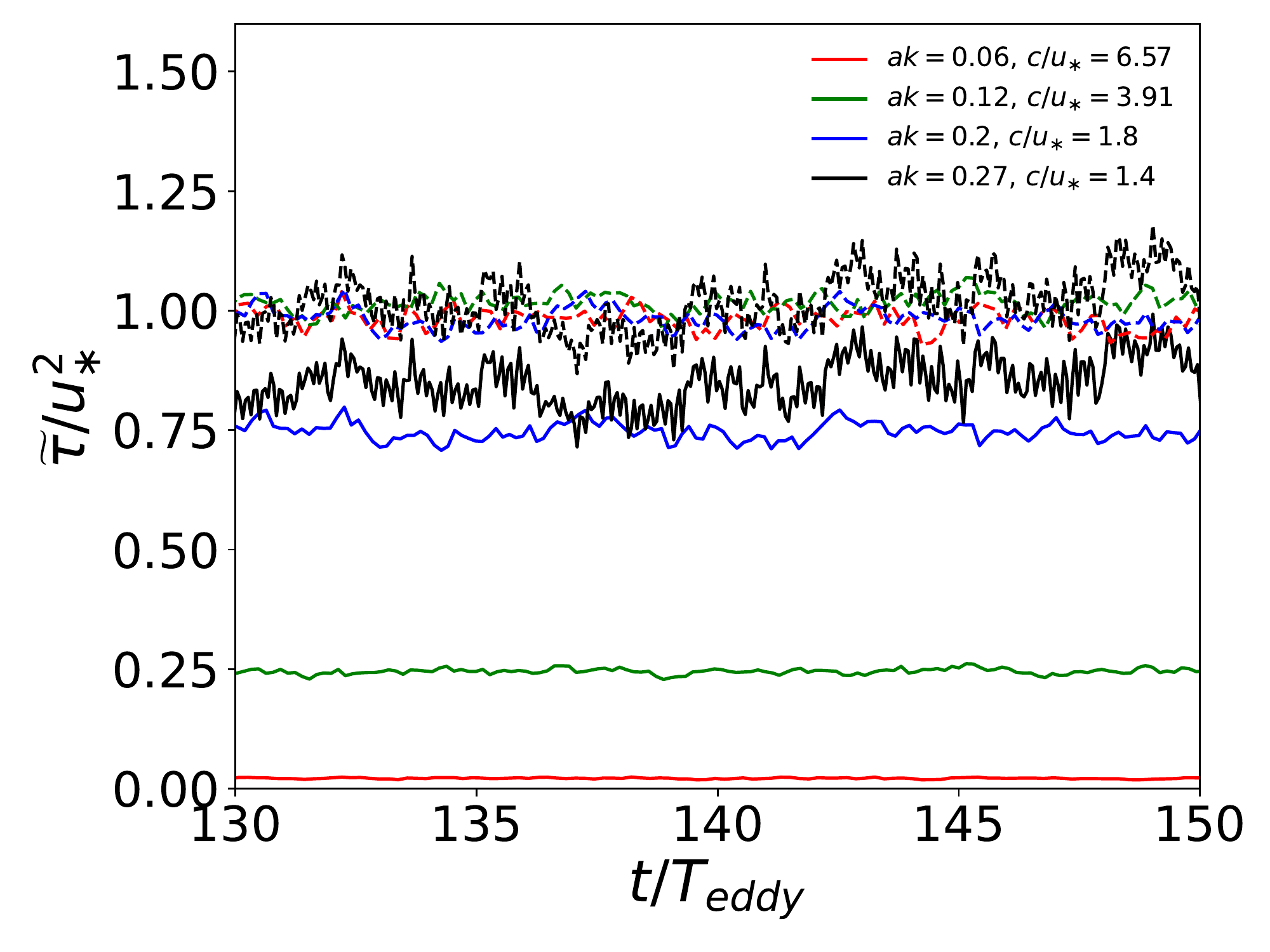}
    \caption{ Normalized form stress ($\Delta_z F_{d,i}$) and total stress ($\langle\ttau_{i3}^T\rangle$) as a function of non-dimensional time for different wave steepnesses, and wave ages. The solid lines represent the form stress, and the corresponding dashed lines represent the total stress for $ak = 0.2$, $c/u_{\ast} = 1.8$; 
    $ak = 0.12$, $c/u_{\ast} = 3.91$;
    and $ak = 0.06$, $c/u_{\ast} = 6.57$.
    The contribution of the form stress increases as a function of wave steepness.  }
    \label{fig:stress_part}
\end{figure}

The averaged total shear stress consists of the averaged resolved Reynolds stress $\langle \tu^{\prime} \tw^{\prime} \rangle$, where $\tu^{\prime} = \tu - \overline{\tu}$ is the fluctuating part of the resolved LES velocity obtained by subtracting the time averaged part of the velocity, and the mean subfilter shear  stress $\langle{\ttau}_{xz}\rangle$. The   normalized resolved and subfilter shear stresses from the four validation cases are shown in Figure \ref{fig:uv_LES}. For each case, the normalized resolved stress shown in Figure \ref{fig:uv_LES}a increases linearly from the top boundary towards the surface and is the dominant contribution to the total stress. The mean normalized  subfilter stress depicted in Figure \ref{fig:uv_LES}b is negligible far from the surface and starts to increase rapidly below $z/H = 0.1$. The total normalized stress shown in Figure \ref{fig:uv_LES}c is similar to a pressure gradient driven turbulent boundary layer. The total normalized shear stress does not reach $-1$ at the surface \CDS{as it does not include the model form stress.}  \CDS{At the surface, the total stress $ \widetilde{\tau}\vert_{wall} = \tau_{xz} +  \rho^{-1}{F}_d\Delta z$ is the sum of the form stress imposed by the model and the subfilter stress (from MOST).} Figure \ref{fig:stress_part} shows the partition of the total surface stress as a function of time for the different cases and  verifies that the wave stress and the subfilter stress sums to the imposed non-dimensional pressure gradient for each case considered.


\subsubsection{Model Form Drag}
\begin{figure}
    \centering
    \includegraphics[width = 0.8\textwidth]{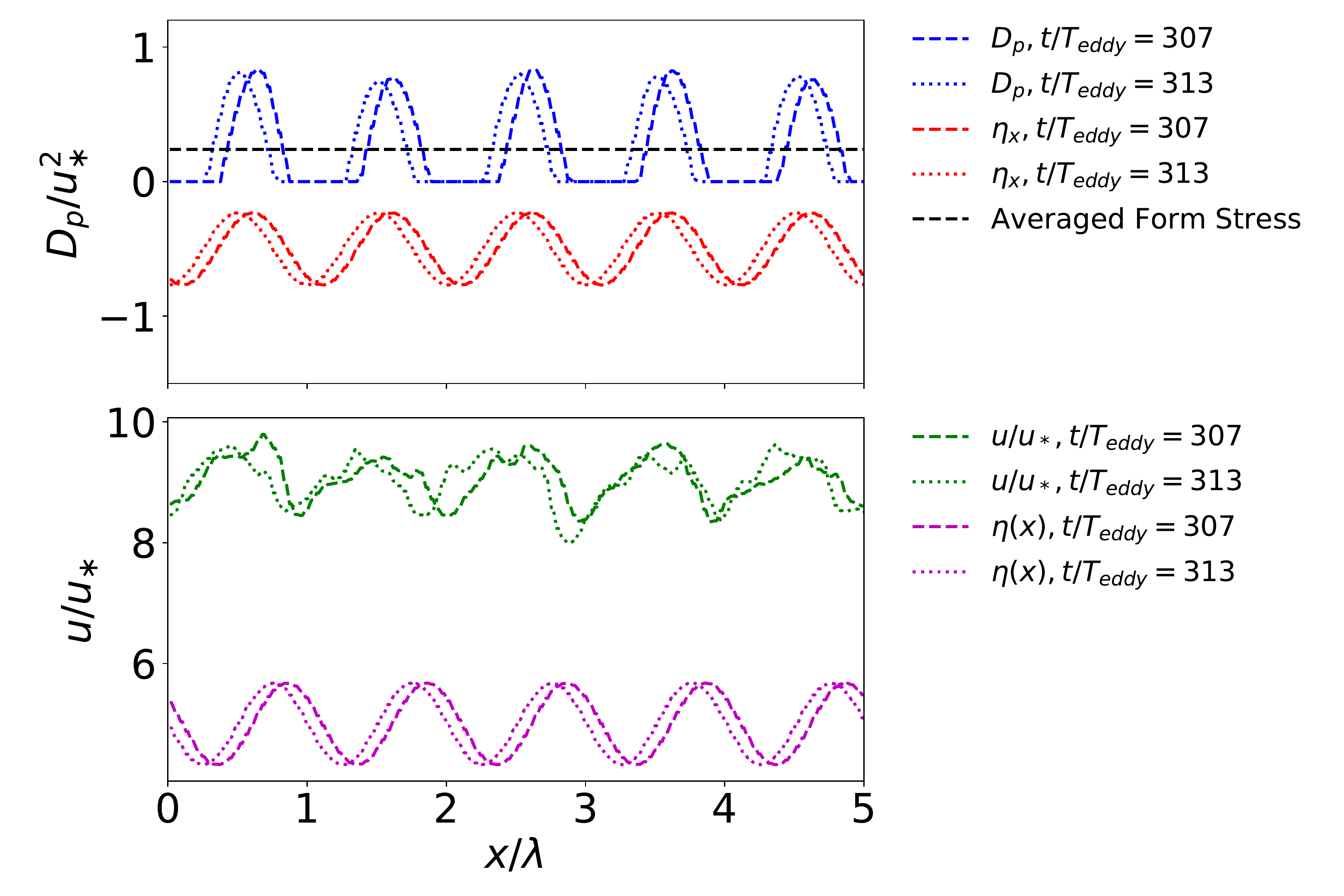}
    \caption{ {The top panel depicts the scaled wave surface gradient distribution and the spanwise averaged normalized form stress $D_p$ at two time instances. Additionally shown in the top panel is the time averaged normalized form stress. The wave drag force is correlated to the wave slope and has a phase difference of $\pi/2$ with respect to the wave height distribution.
    The bottom panel depicts the spanwise averaged normalized streamwise velocity at the wave surface as a function of horizontal distance at two time instances. Also shown in the bottom panel is the corresponding wave height distribution (magenta lines). The velocity shows the characteristic acceleration on the leeward  side of the wave with respect to the wave trough followed by deceleration. The data corresponds to the case with $ak = 0.12$ and $c/u_* = 3.91$.} }
    \label{fig:Force_and_vel}
\end{figure}

To understand the effects of the wave drag model in the LES, the spanwise-averaged velocity and the wave  {form stress calculated from the drag force} applied at the first grid point are analyzed in Figure \ref{fig:Force_and_vel} for $ak = 0.12$. The form  {stress} for phase-resolved simulations is obtained from the resolved pressure $ \widetilde{p}\partial\widetilde{\eta}/\partial x$.
In the current LES, as the waves are unresolved in the vertical direction, the model form stress $D_p$ depends on the incoming momentum flux $M_{wave}$ and the drag coefficient for the wind wave transfer. \CDS{The stress is calculated using the drag force per unit volume ${F}_d$ defined in Equation $\ref{eqn:drag_eqn_final}$ multiplied by the vertical grid spacing $\Delta_z$:}
\begin{equation}\label{eqn:Dp}
    D_p = \CDS{\frac{C_D}{\rho}}\frac{M_{wave}}{(\Delta_x\Delta_y)} = \frac{1}{\rho}|{\bm{F}}_d|\Delta_z=\frac{P}{1 + Q(ak)^2}(ak)\tu U^{\Delta} \left(\widehat{n}_{u,k}\cdot \frac{\partial \widetilde{\eta}}{\partial x_k}\right) {H\left\{\widehat{n}_{u,k}\cdot \frac{\partial \widetilde{\eta}}{\partial x_k}\right\}}.
\end{equation}
 The top panel depicts the spanwise averaged model form stress for two time instants. The drag force is correlated to the surface gradient of the wave height distribution and is out of phase with the wave surface (shown in the bottom panel) by a factor of $\pi/2$. The magnitude of the peaks of the drag force shows temporal dependence due to the variability of the turbulent velocity at the first grid point (shown in the bottom panel). Additionally shown in the top panel is the time and space averaged value of the normalized form stress.  There is considerable variation of the drag applied along the wave surface with respect to the mean. This effect is also observed in experiments and phase-resolved simulations \citep{Hara2015,Sullivan2018a,Husain2019,Buckley2020,yousefi_veron_buckley_2020}. The bottom panel of Figure \ref{fig:Force_and_vel} depicts the spanwise average of the streamwise velocity at the wave surface (first grid point) for two distinct time steps. The wave height distribution is also depicted for reference. The velocity accelerates in the upstream side of the surface peak where correspondingly the force applied by the model (top panel of Figure \ref{fig:Force_and_vel}) is zero and then decelerates as the flow encounters the next wave surface corresponding to a maximum in the applied force. Phase-averaged models purely apply the averaged drag and would not capture this variation in the velocity and drag. This variation is important for calculating hydrodynamic forces for offshore structures.

\begin{figure}
    \centering
    \includegraphics[width=\textwidth]{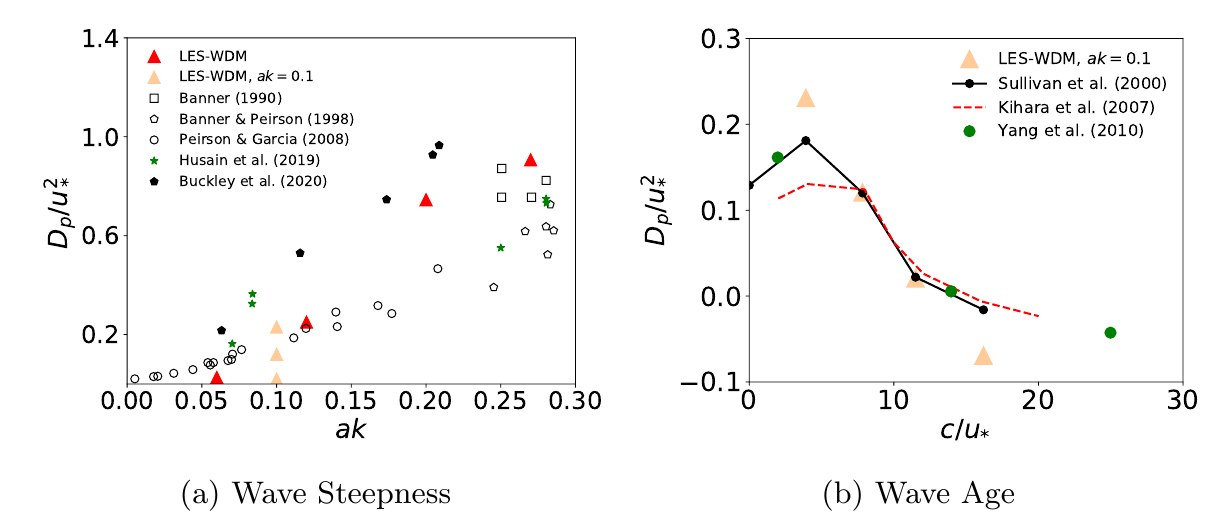}
    \caption{ Normalized form stress as a function of a) wave steepness and b) wave age. In the left panel, the symbols correspond to simulation results from the  LES-WDM (current work), LES from \citet{Husain2019}, and experimental data from  
    \citet{Banner1990}, 
    \citet{banner_peirson_1998}, 
    \citet{Peirson2008}, and \citet{Buckley2020}. In the left panel, apart from the LES-WDM validation cases, the form stress from the LES at $ak = 0.1$ and  $c/u_* = 4.2,\ 7.84$ and $11.6$ are shown. 
    In the right panel, the LES-WDM at different wave ages $c/u_* = 4.2,\ 7.84$ and $11.6$ and $ak=0.1$ is compared to DNS data from \citet{Sullivan2000}, \citet{Kihara2007}, and \citet{Yang2010}.
    The model shows good agreement with the existing experimental and numerical data.}
    \label{fig:form_drag}
\end{figure}

Figure \ref{fig:form_drag}a highlights the  average normalized form stress defined in Equation \ref{eqn:Dp} as a function of wave steepness. The LES-WDM shows good agreement with the experimental data from \citet{Banner1990}, \citet{banner_peirson_1998}, and \citet{Peirson2008} and accurately predicts the change of form stress for different steepness values considered here. The stresses from \citet{Buckley2020} are higher than the other reported experimental values, but the variation of form stress as a function of wave-steepness shows similar qualitative trend. The experimental scatter is in part due to the uncertainties in the measurements. Note that the form stress calculated here is the stress due to the wave drag model acting on the first grid point. The effect of changing wave age for a fixed wave steepness reduces the form stress as the faster waves provide smaller resistance to the airflow above.
The normalized form stress as a function of the wave age $c/u_{\ast}$ is shown in Figure \ref{fig:form_drag}b. The simulations are run at a fixed steepness, and the grid is set such that $a = 0.95\Delta z/2$. The LES-WDM accurately predicts the form stress from the DNS datasets of \citet{Sullivan2000}, \citet{Kihara2007}, and \citet{Yang2010} as a function of wave age for $c/u_{\ast} < 15$. For very fast waves ($\tuu_r <0$ at the first grid point), the model predicts a negative form stress based on Equation (\ref{eqn:drag_eqn_final}), following similar qualitative trends to the DNS, but the magnitude of the negative form stress is too large. The transfer of momentum is from the waves to the wind, a scenario common during conditions of swell, where fast moving waves (with $c/u_* > 25$) generated from distant systems transfer energy to the slower wind. Future studies could work towards developing a model for fast moving waves.

\section{Summary and Conclusions}\label{sec:Conclusions}

Characterizing the momentum transfer between wind and waves is critical in numerous atmospheric applications. Wave phase-resolving simulations through immersed boundary or terrain following approaches come with a high computational cost for simulating atmospheric flows. On the other extreme, wave phase-averaged approaches with an effective roughness length are not sufficiently generalizable to a broad range of conditions. In this work a wave-based drag model applicable to wall-modeled Large Eddy Simulations is proposed.  The model quantifies the momentum transfer between the wind and the wave-field through a canopy stress approach. The surface-gradient based approach of \citet{Anderson2010} is extended to moving boundaries and the formulation adapted for the wind-wave problem. The drag coefficient is found to be proportional to the wave steepness  to give the correct scaling for the form stress.

The model is applied and tested on various cases of turbulent boundary layer flow over idealized water waves of different wave steepnesses and wave ages. \CDS{The "phase-aware" nature of the model is well captured by the instantaneous velocity profiles and the phase-averaged streamwise wave fluctuation velocity.} The predicted mean velocity profiles show good agreement with data from experiments. The form stress predicted by the model is in good agreement with existing experimental data and shows the correct trend with both wave age and wave steepness. For faster waves, the model is capable of predicting a  negative form stress, where the momentum is transferred from the wave-field to the wind.  
The model showed   generalizability over the parameter space studied when compared to phase-averaged roughness models and requires no tuning parameters or empirical constants. The inherent spatial and temporal variability associated with the the drag force imposed by the waves on the airflow can also be well represented. \CDS{Vertical wave induced motions are not accounted for explicitly in the model; however, both the wave-induced stress and the pressure stress are captured implicitly.
Incorporating the effects of wave-induced motions to the vertical velocity is important for swell conditions (where the waves move faster than the near-surface wind velocity) and  will be the focus of future studies.}

The framework proposed here to incorporate wave effects can serve as a low-cost accurate LES framework to simulate airflow in the marine atmospheric boundary layer with a significant reduction in computational cost compared to phase-resolved approaches.  {The coarse resolution used here allows the accurate prediction of the mean velocity profiles and wave stress with an $O(10^4)$ reduction in computational cost based on grid resolution.}  {The generalizability and ease of implementation of the wave drag model allows scalability to large domains and could be useful in the context of offshore wind farm (fixed bottom and floating turbines) simulations, marine atmospheric boundary layer simulations, and ocean mixed layer simulations.}

\acknowledgments
The authors gratefully acknowledge financial support from the Princeton University Andlinger Center for Energy and the Environment and High Meadows Environmental Institute. The simulations presented in this article were performed on computational resources supported by the Princeton Institute for Computational Science and Engineering (PICSciE) and the Office of Information Technology's High Performance Computing Center and Visualization Laboratory at Princeton University.
%
%
\datastatement
Data can be made available upon request.

%






%
%
%
\bibliographystyle{ametsoc2014}
\bibliography{references,references_2}

\end{document}